\newif\ifarxiv
\arxivtrue
\documentclass[letterpaper]{article} 
\usepackage{aaai2026}  
\usepackage{times}  
\usepackage{helvet}  
\usepackage{courier}  
\usepackage[hyphens]{url}  
\usepackage{graphicx} 
\usepackage{natbib}  
\usepackage{caption} 
\usepackage{algorithm}
\usepackage{algorithmic}
\usepackage[utf8]{inputenc} 
\usepackage[T1]{fontenc}    
\ifarxiv
  \usepackage{hyperref}
\fi
\usepackage{url}            
\usepackage{booktabs}       
\usepackage{amsfonts}       
\usepackage{graphicx}       
\usepackage{nicefrac}       
\usepackage{microtype}      
\usepackage{xcolor}         
\usepackage{tabularx}
\usepackage[most]{tcolorbox}
\usepackage{amsmath}
\usepackage{colortbl} %
\definecolor{grayrow}{gray}{0.92}
\usepackage{caption}
\usepackage{CJKutf8} 
\usepackage{pdfpages}
\usepackage{fontawesome5}
\usepackage{subcaption}
\usepackage{newfloat}
\usepackage{listings}
\DeclareCaptionStyle{ruled}{labelfont=normalfont,labelsep=colon,strut=off} 
\floatstyle{ruled}
\newfloat{listing}{tb}{lst}{}
\floatname{listing}{Listing}
\title{NVSpeech: An Integrated and Scalable Pipeline for Human-Like Speech Modeling with Paralinguistic Vocalizations}
\author {
    Huan Liao\textsuperscript{\rm 1}\thanks{Equal contribution.},
    Qinke Ni\textsuperscript{\rm 1}\footnotemark[1],
    Yuancheng Wang\textsuperscript{\rm 1},
    Yiheng Lu\textsuperscript{\rm 1},\\
    Haoyue Zhan\textsuperscript{\rm 2},
    Pengyuan Xie\textsuperscript{\rm 2},
    Qiang Zhang\textsuperscript{\rm 2},
    Zhizheng Wu\textsuperscript{\rm 1},
}
\affiliations {
    \textsuperscript{\rm 1}The Chinese University of Hong Kong, Shenzhen\\
    \textsuperscript{\rm 2}Guangzhou Quwan Network Technology\\
}

\usepackage{bibentry}

\begin{document}
\begin{CJK*}{UTF8}{gbsn} 
\maketitle

\begin{abstract}
Paralinguistic vocalizations—including non-verbal sounds like laughter and breathing, as well as lexicalized interjections such as “uhm” and “oh”—are integral to natural spoken communication. Despite their importance in conveying affect, intent, and interactional cues, such cues remain largely overlooked in conventional automatic speech recognition (ASR) and text-to-speech (TTS) systems.
We present \textbf{NVSpeech}, an integrated and scalable pipeline that bridges the recognition and synthesis of paralinguistic vocalizations, encompassing dataset construction, ASR modeling, and controllable TTS.
(1)~We introduce a manually annotated dataset of 48,430 human-spoken utterances with 18 word-level paralinguistic categories.
(2)~We develop the \textit{paralinguistic-aware ASR model}, which treats paralinguistic cues as inline decodable tokens (e.g., “You’re so funny [Laughter]”), enabling joint lexical and non-verbal transcription. This model is then used to automatically annotate a large corpus, the first large-scale Chinese dataset of 174,179 utterances (573 hours) with word-level alignment and paralingustic cues.
(3)~We finetune zero-shot TTS models on both human- and auto-labeled data to enable explicit control over paralinguistic vocalizations, allowing context-aware insertion at arbitrary token positions for human-like speech synthesis.
By unifying the recognition and generation of paralinguistic vocalizations, \textbf{NVSpeech} offers the first open, large-scale, word-level annotated pipeline for expressive speech modeling in Mandarin, integrating recognition and synthesis in a scalable and controllable manner. Dataset and audio demos are available at \href{https://nvspeech170k.github.io/}{https://nvspeech170k.github.io/}.

\end{abstract}

\section{Introduction}
Paralinguistic vocalizations—such as nonverbal vocalizations (NVVs) like laughter and breathing, as well as lexicalized interjections like “uhm” and “oh”—are widely present in spontaneous speech~\cite{tseng2003taxonomy}. These cues, especially interjections, encode affect, intent, and speaker state beyond literal lexical contents, often via distinctive prosody, enhancing expressivity and ensuring social appropriateness~\cite{loy2017effects, kidd2011toddlers, ward2006non}. However, paralinguistic vocalizations, particularly NVVs, are often discarded as noise in conventional speech processing pipelines due to the lack of annotated fine-grained, word-level alignment data.

Recent advances in automatic speech recognition (ASR) and text-to-speech (TTS) systems have led to impressive gains in transcription accuracy and speech naturalness. However, traditional ASR focuses solely on lexical contents, overlooking paralinguistic cues crucial for understanding spontaneous communication~\cite{lee2003designing, gao2305funasr}. Similarly, recent TTS systems~\cite{an2024funaudiollm, du2024cosyvoice, guo2024fireredtts} enable instruction-based paralinguistic control, but typically rely on closed-source datasets with limited behavioral diversity, lacking transparent and fine-grained supervision for modeling natural paralinguistic vocalizations.

\begin{figure*}[t!]
\centering
\includegraphics[width=1\linewidth]{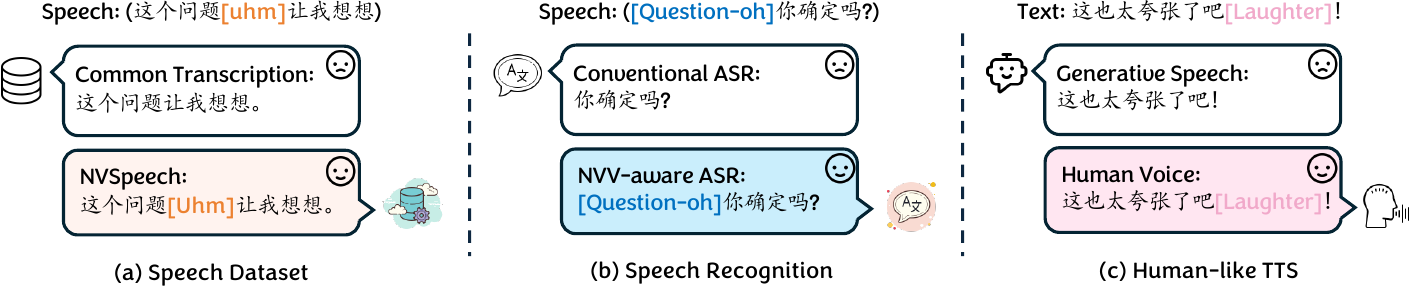}
\caption{The gap between conventional speech processing systems and paralinguistic-aware modeling.
\textbf{(a)} Common speech datasets omit paralinguistic vocalizations, while NVSpeech provides word-level annotations.
\textbf{(b)} Conventional ASR ignores such cues; our paralinguistic-aware ASR jointly transcribes lexical and non-lexical content.
\textbf{(c)} Standard TTS generates only text-based speech, while our TTS supports explicit insertion of paralinguistic vocalizations for human-like synthesis.
}
\label{fig:motivation}
\end{figure*}

In real-world speech, verbal and non-verbal cues are inherently intertwined. To capture human-like communication, ASR and TTS systems should move beyond lexical content to model paralinguistic signals in a temporally aligned and semantically coherent manner. In Figure~\ref{fig:motivation}, we identify three critical gaps in current speech modeling:
(a)~Most existing speech datasets lack word-level annotations of paralinguistic vocalizations, limiting supervision and evaluation of expressive models.
(b)~Conventional ASR systems omit such cues, hindering their application in tasks requiring human-like understanding and interaction\cite{lee2003designing, gao2305funasr}.
(c)~Current TTS models fail to synthesize paralinguistic vocalizations with explicit, token-level control, resulting in speech that lacks spontaneity and expressivity.

In tonal languages like Mandarin, paralinguistic cues interact closely with tone and prosody shaping affect, discourse, and speaker intent. Interjections, hesitations, and NVVs play key roles in marking turn-taking, signaling uncertainty or emphasis in spontaneous speech. Without fine-grained, aligned annotations, models struggle to capture these nuances, resulting in unnatural synthesis and fragile speaker-state recognition. Existing Chinese corpora often overlook or coarsely label such cues, hindering both supervision and evaluation of human-like speech understanding and generation.

To bridge these gaps, we introduce \textbf{NVSpeech}, an integrated and scalable pipeline for recognition and synthesis of paralinguistic vocalizations in Chinese speech. It centers on a large-scale, word-level annotated corpus with 18 types of paralinguistic vocalizations—ranging from NVVs such as \texttt{[Laughter]} and \texttt{[Cough]}, to prosodic and attitudinal interjections like \texttt{[Confirmation-en]}, \texttt{[Question-ah]}, and discourse markers \texttt{[Uhm]}. These annotations provide word-alignment and broad coverage, making NVSpeech the first corpus to support fine-grained modeling of both lexical and paralinguistic cues in Mandarin.

We implement \textbf{NVSpeech} in three stages:

\textbf{(1) Manual annotation.}
We collect a high-quality subset of 48,430 utterances from real-world, human-spoken recordings. Each utterance is manually annotated at the word level with paralinguistic vocalization labels, spanning 18 categories to capture expressive behaviors beyond lexical content. This annotated subset serves as the foundation for training our paralinguistic-aware models.

\textbf{(2) Scalable labeling via paralinguistic-aware ASR.}
Using the manually annotated data, we train the first \textit{paralinguistic-aware ASR model} capable of transcribing both lexical content and inline paralinguistic vocalizations, as illustrated in Figure~\ref{fig:motivation}(b). We apply this model to a large unlabeled speech corpus, including data from miHoYo and Emilia~\cite{he2025emilia}, resulting in an auto-labeled dataset of 174,179 utterances. This process dramatically scales up annotation coverage while significantly reducing human labeling cost.

\textbf{(3) Expressive TTS modeling.}
To validate the utility of \textbf{NVSpeech}, we finetune zero-shot TTS models on both manually and automatically labeled data~\cite{tan2021survey}. The model enables explicit control over paralinguistic vocalizations, allowing expressive and context-aware insertion at arbitrary word positions—achieving controllable synthesis beyond the expressivity of conventional TTS systems.

\noindent \textbf{Our main contributions are:}

(1) We develop the first \textit{paralinguistic-aware ASR model} that jointly transcribes lexical content and paralinguistic vocalizations with word-level alignment, enabling structured modeling of expressive speech beyond conventional ASR.

(2) We present \textbf{NVSpeech}, an integrated and scalable pipeline that integrates data, ASR, and TTS,
centered on a large-scale corpus with word-level annotations for 18 categories of paralinguistic vocalizations. The corpus includes both manually annotated and auto-labeled subsets, totaling 573.4 hours, and supports both recognition and generation of human-like vocal behaviors with explicit controllability.

(3) We conduct comprehensive benchmarking on paralinguistic tagging, ASR, and zero-shot TTS tasks, demonstrating that \textbf{NVSpeech} enables controllable paralinguistic cues insertion and improves naturalness of synthesized speech.

\section{Related Works}

\subsection{Paralinguistic Event Recognition}
Spontaneous speech ASR has advanced with DNN-based approaches. Due to high annotation costs, Mandarin lacks richly transcribed corpora comparable to English’s Switchboard and Fisher; consequently, most studies ignore non-speech cues (e.g., laughter, breathing) and seldom incorporate discourse markers (e.g., ‘uhm’, ‘oh’) into decoding.

Early works~\cite{gupta2016detecting},\cite{rennie2022model} leveraged prosodic features and MFCCs with classifiers like HMMs to detect laughter and fillers at the frame level. While effective, these models relied on handcrafted features and limited data. Pretrained Audio Event Detection (AED) models like PANNs~\cite{kong2020panns} and BEATs~\cite{chen2022beats}, can detect sound events and provide general audio representations. However, these resources lack word-level alignment, are not optimized for conversational speech, and typically cover only sound events, excluding linguistic interjections like ‘uhm’ or ‘oh.’

SenseVoice~\cite{an2024funaudiollm}augments voice understanding with auxiliary tasks via task-specific tokens and pseudo-labeled data; however, it treats event detection as decoupled from speech recognition and does not explicitly model interactions between verbal and nonverbal components.

\begin{table*}[t!]
  \centering
  \small
  \begin{tabular}{p{6.4cm} c c c c c c c}
    \toprule
    Dataset & \#Utterances & Total (h)  & \#Classes & \#Speaker & Annot. Level & Lang & Avail. \\
    \midrule
    Sinica MCDC8~\cite{tseng2013lexical} & 30 & 25.6 & 32 & 16 & Segment & CN & Public\\
    SMC~\cite{polychroniou2014sspnet} & - & 12.68 & 5 & 120 & Segment & EN & Public\\
    SVC~\cite{salamin2013automatic} & 2,763 & 8.4 & 6 & 120 & Segment & EN & Public\\
    RAMC~\cite{yang2022open} & 219,325 & 180 & 3 & 663 & Segment & CN & Public\\
    NVV & 70,000 & 56.7 & 16 & 1,419 & Sentence & EN & Public\\
    VocalSound~\cite{gong2022vocalsound} & 21,024 & - & 6 & 3,365 & Sentence & Multi & Public\\
    Nonspeech7k~\cite{rashid2023nonspeech7k} & 7,014 & 6.75 & 7 & - & Sentence & EN & Public\\
    MCDC~\cite{deng2023toward} & 7,014 & 6.75 & 7 & - & Sentence & EN & Public\\
    EXPRESSO~\cite{nguyen2023expresso} & 11,615 & 47 & - & 4 & Sentence & EN & Public\\ 
    DisfluencySpeech~\cite{wang2024disfluencyspeech} & 5,000 & 9.49 & 15 & 1 & Word & EN & Public \\
    \midrule
    NVSpeech\textsubscript{human} & 48,430  & 76 & 18 & 1,578 & Word & CN & Private\\
    NVSpeech & 174,179  & 573.4 & 18 & >1,964 & Word & CN & Public\\
    \bottomrule
  \end{tabular}
  \caption{Comparison of Paralinguistic Datasets}
  \label{tab:para-datasets}
\end{table*}

\subsection{Datasets with Paralinguistic Label}
Several datasets have been developed to support the study of paralinguistic vocalizations, as summarized in Table~\ref{tab:para-datasets}. Early corpora such as the SSPNet Mobile Corpus (SMC)~\cite{polychroniou2014sspnet} and SSPNet Vocalization Corpus (SVC)~\cite{salamin2013automatic} provide manually segmented annotations of laughter and fillers (e.g., uhm, eh) in phone call-mediated settings. Similarly, DisfluencySpeech~\cite{wang2024disfluencyspeech}, a English speech dataset, offers word-level annotations of fillers, discourse markers like ``you know'' and ``well'', and NVVs (e.g., sigh, laughter) in clean monologue recordings.

More recent efforts focus on broader coverage and speaker diversity. VocalSound~\cite{gong2022vocalsound} and Nonspeech7k~\cite{rashid2023nonspeech7k} compile sentence-level clips across NVVs classes, while the NVV\footnote{\url{https://github.com/deeplyinc/Nonverbal-Vocalization-Dataset}} dataset covers 16 types of human nonverbal sound, as well as less commonly annotated events such as teeth-chattering. The EXPRESSO corpus~\cite{nguyen2023expresso} emphasizes expressive and improvised styles, capturing spontaneous non-verbals and providing benchmark tools for expressive synthesis. Meanwhile, RAMC~\cite{yang2022open} includes categories like laughter and crying but is limited in event diversity.

Despite these efforts, most existing datasets suffer from several limitations:
(1) lack of word-level alignment between lexical and non-verbal content, hindering precise modeling and in-context understanding or synthesis;
(2) limited speaker diversity or constrained recording scenarios, with scarce Chinese data;
(3) existing jointly annotated corpora either cover few paralingustic types or only sentence-level labels. 

\subsection{Human-like TTS with Paralinguistic Vocalization}
Recent advances in human-like TTS aim to incorporate paralinguistic vocalizations to improve speech expressiveness. NSV-TTS~\cite{zhang2023nsv} jointly models speech and NVVs using a hybrid representation of phonemes and unsupervised linguistic units (ULUs), enabling zero-shot synthesis of events such as cough and cries.
Several approaches introduce disfluency and paralinguistic behaviors through text-based control. CosyVoice-Instruct\cite{an2024funaudiollm} and FireRedTTS\cite{guo2024fireredtts} allow users to insert NVVs via text tokens or embeddings, supporting behaviors like repetition and emphasis. While these methods offer flexible control, both rely on private datasets.

Chaudhury et al.~\cite{chaudhury2024humane} use a language model to insert disfluency markers with a rule-based TTS backend, but suffer from poor interpretability, no personalization, and static mappings unsuited to dynamic disfluent speech. EmoCtrl-TTS~\cite{wu2024laugh} employs a flow-matching model trained on pseudo-labeled embeddings for emotion and NVVs. However, its NVVs modeling is restricted to laughter and cry, and it requires external NVV prompt embeddings during inference, which limits flexibility and scalability. Similarly, ELaTE~\cite{kanda2024making} focuses solely on laughter via frame-level conditioning.

While recent TTS have explored integrating paralinguistic vocalizations to enhance expressiveness, they suffer from several limitations. Many rely on small-scale or private datasets with limited NVVs coverage, their generation strategies often lack word-level alignment or require external embeddings or detectors during inference. Text-controlled systems like FireRedTTS offer flexibility but depend on static mappings, heuristic rules, or exhibit poor interpretability. 
In contrast, our approach leverages a publicly available word-level annotated dataset covering diverse paralinguistic vocalizations, enabling token-level paralinguistic vocalizations insertion for scalable, controllable, and natural TTS. 

\begin{figure*}[h!]
    \centering
    \includegraphics[width=1\linewidth]{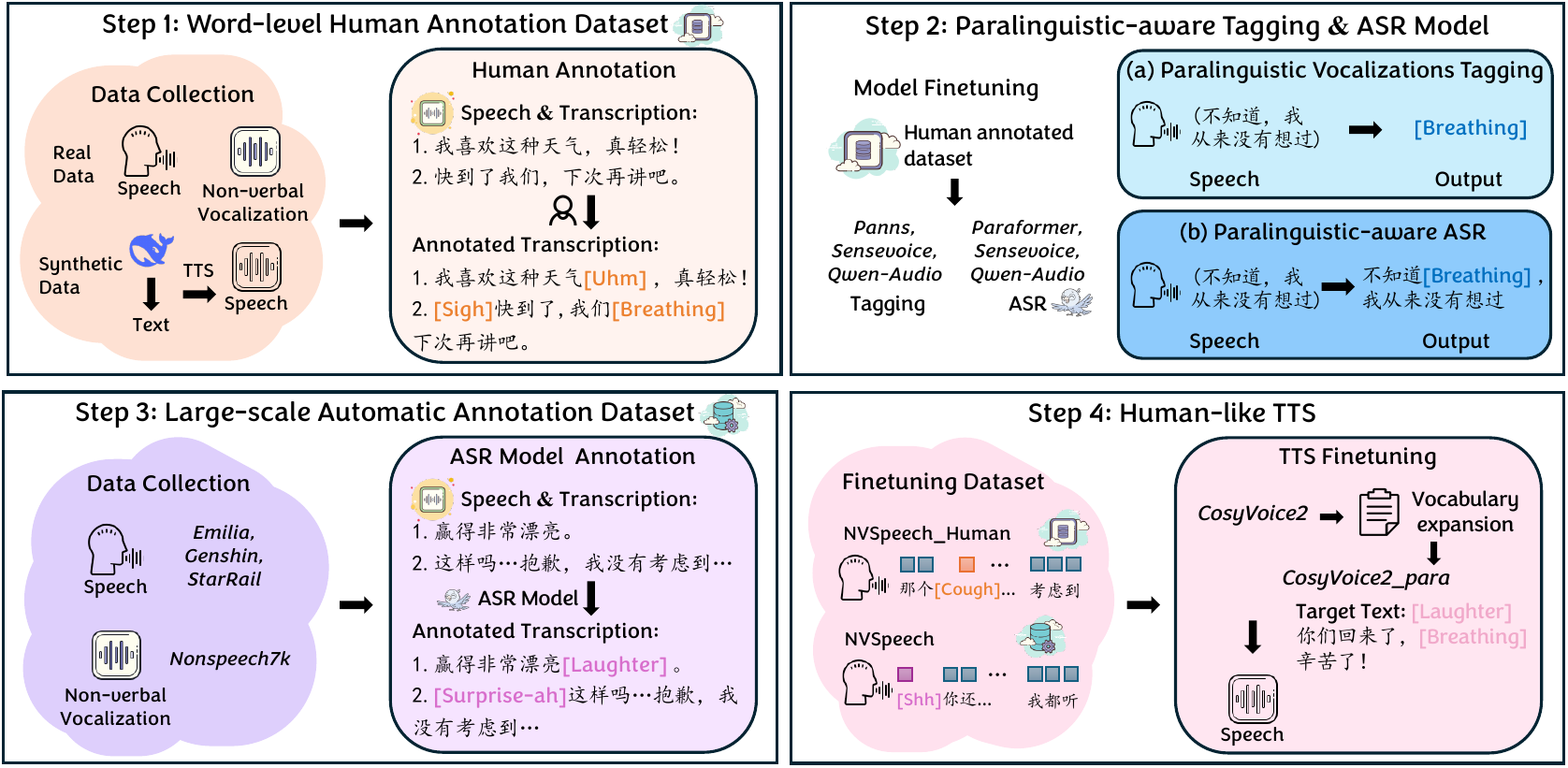}
    \caption{Overview of our paralinguistic-aware speech recognition and generation pipeline.(1) A word-level human annotated dataset of verbal and non-verbal vocalizations is first constructed. (2) A paralinguistic-aware ASR model is trained to jointly transcribe verbal and non-verbal content. (3) This model is used to automatically annotate large-scale unlabeled speech. (4) The expanded dataset enables training a controllable and expressive TTS system that explicitly renders paralinguistic cues.}
    \vspace{-0.5cm}
    \label{fig:enter-label}
\end{figure*}
\section{Paralinguistic Speech Recognition Model}

\subsection{Paralinguistic Tagging}
\paragraph{Experimental Setup}
We formulate paralinguistic tagging as a multi-label classification problem, where each utterance is assigned one or more paralinguistic vocalization tags $y \in [0,1]^C$ from a predefined vocabulary of size $C$. Given input audio $x \in \mathbb{R}^{T \times F}$, the model predicts $\hat{y} \in [0,1]^C$. Training are conducted on the human-annotated subset of \textbf{NVSpeech}, with samples in the form $(x, y)$ where $y$ is the sentence-level ground-truth tag vector. The training objective is the binary cross-entropy loss:
\begin{equation}
\mathcal{L}_{\text{BCE}} = -\sum_{c=1}^{C} \left[ y_c \log \hat{y}_c + (1 - y_c) \log (1 - \hat{y}_c) \right]
\end{equation}
We finetune three audio tagging models on the training split.

\paragraph{Evaluation}
We evaluate on the held-out test split of the human-labeled subset, where each sample is $(x, y)$ with $y$ representing the sentence-level ground-truth tag vector. Performance is reported using standard multi-label classification metrics: \textit{Precision}, \textit{Recall}, and \textit{F1 score}.

\paragraph{Baseline Models} We evaluate three baselines: (1) PANNs~\cite{kong2020panns}: A CNN-based audio tagging model pretrained on AudioSet~\cite{jort_audioset_2017}. We finetune the \texttt{Wavegram\_Logmel\_Cnn14}, which extracts frame-level features and applies pooling for utterance-level multi-label prediction. (2) SenseVoice-Small~\cite{an2024funaudiollm}: A speech foundation model pretrained with pseudo-labeled audio events. We extend its single-label prediction to multi-label tagging by allowing up to five tags per utterance—reflecting the maximum in our dataset—and padding with \texttt{[None]} when fewer events are present. (3) Qwen-Audio~\cite{chu2023qwen}: A Whisper-style encoder + Qwen-7B decoder trained on over 30 audio-text tasks. We finetune it using instruction-style prompts and sentence-level tag supervision. The prompt format is illustrated in Table ~\ref{tab:qwen-prompt}.

\begin{table}[h!]
\centering
\small
\begin{tcolorbox}[colback=gray!5, colframe=black, width=\columnwidth, boxrule=1pt]
\textbf{Prompt for Paralinguistic Tagging:} \\
\texttt{<audio>} Given an audio clip, identify the paralinguistic event from the following EVENT LABEL SET: {\texttt{\{[Breathing], [Crying], [Laughter], ..., [Shh]\}}} \\
MUST follow this template: \\
The paralinguistic vocalizations detected are: \texttt{\{[EVENT 1], [EVENT 2], ..., [EVENT N]\}} \\
OR the paralinguistic vocalization detected in the audio clip is \texttt{[None]}.
\end{tcolorbox}
\captionof{table}{Instruction prompt used for Qwen-Audio.}
\label{tab:qwen-prompt}
\end{table}

\paragraph{Results}
As shown in Table~\ref{tab:tagging-table}, SenseVoice achieves the best overall F1 score (0.73) on the NVSpeech\_test set, demonstrating the benefit of pseudo-labeled data and ASR-aware training. PANNs performs competitively with strong precision, confirming its strength in audio event detection. 

\begin{table}[h!]
    \centering
    \small
    \begin{tabular}{p{1.8cm}p{1.2cm}ccc}
        \toprule
        \textbf{Model} & \textbf{Arch.} & Precision$\uparrow$ & Recall$\uparrow$ & F1-score$\uparrow$ \\
        \midrule
        PANNs & CNN & 0.84 & 0.65 & 0.72 \\
        SenseVoice & Transformer & \textbf{0.84} & \textbf{0.67} & \textbf{0.73} \\
        Qwen-Audio & LLM & 0.79 & 0.56 & 0.61 \\
        \bottomrule
    \end{tabular}
    \caption{Paralinguistic Tagging Model Comparison}
    \label{tab:tagging-table}
\end{table}

\begin{figure}[h!]
  \centering
  \begin{minipage}[t]{0.49\textwidth}
    \centering
    \small
    \setlength{\tabcolsep}{1mm}
    \renewcommand{\arraystretch}{1.2}
    \begin{tabular}{lcccc}
      \toprule
      \textbf{Metric} & \textbf{Whisper} & \textbf{Paraformer} & \textbf{SenseVoice} & \textbf{Qwen-Audio} \\
      \rowcolor{grayrow}
      \multicolumn{5}{c}{In-domain Testset} \\
      \midrule
      CER$\downarrow$ & 14.18\% & 4.67\% & \textbf{4.61\%} & 5.47\% \\
      CER$_{\text{w/o para}}$$\downarrow$ & 11.14\% & 2.26\% & \textbf{2.11\%} & 2.62\% \\
      Para Det. Rate$\uparrow$ & 84.8\% & \textbf{96.1\%} & 93.4\% & 94.5\% \\
      F1-score$\uparrow$ & 0.71 & 0.78 & \textbf{0.83} & 0.65 \\
      \rowcolor{grayrow}
      \multicolumn{5}{c}{Open-domain Testset} \\
      \midrule
      CER$\downarrow$ & 19.41\% & 7.81\% & \textbf{3.79\%} & 10.06\% \\
      CER$_{\text{w/o para}}$$\downarrow$ & 16.41\% & 5.30\% & \textbf{3.16\%} & 6.74\% \\
      Para Det. Rate$\uparrow$ & 71.3\% & 74.6\% & \textbf{93.4\%} & 91.0\% \\
      F1-score$\uparrow$ & 0.50 & 0.72 & \textbf{0.85} & 0.54 \\
      \bottomrule
    \end{tabular}
    \captionsetup{type=table}
    \caption{Performance of paralingustic-aware ASR.}
    \label{tab:para-asr-results}
  \end{minipage}
  \hfill
  \begin{minipage}[t]{0.5\textwidth}
    \centering
    \includegraphics[width=0.98\textwidth]{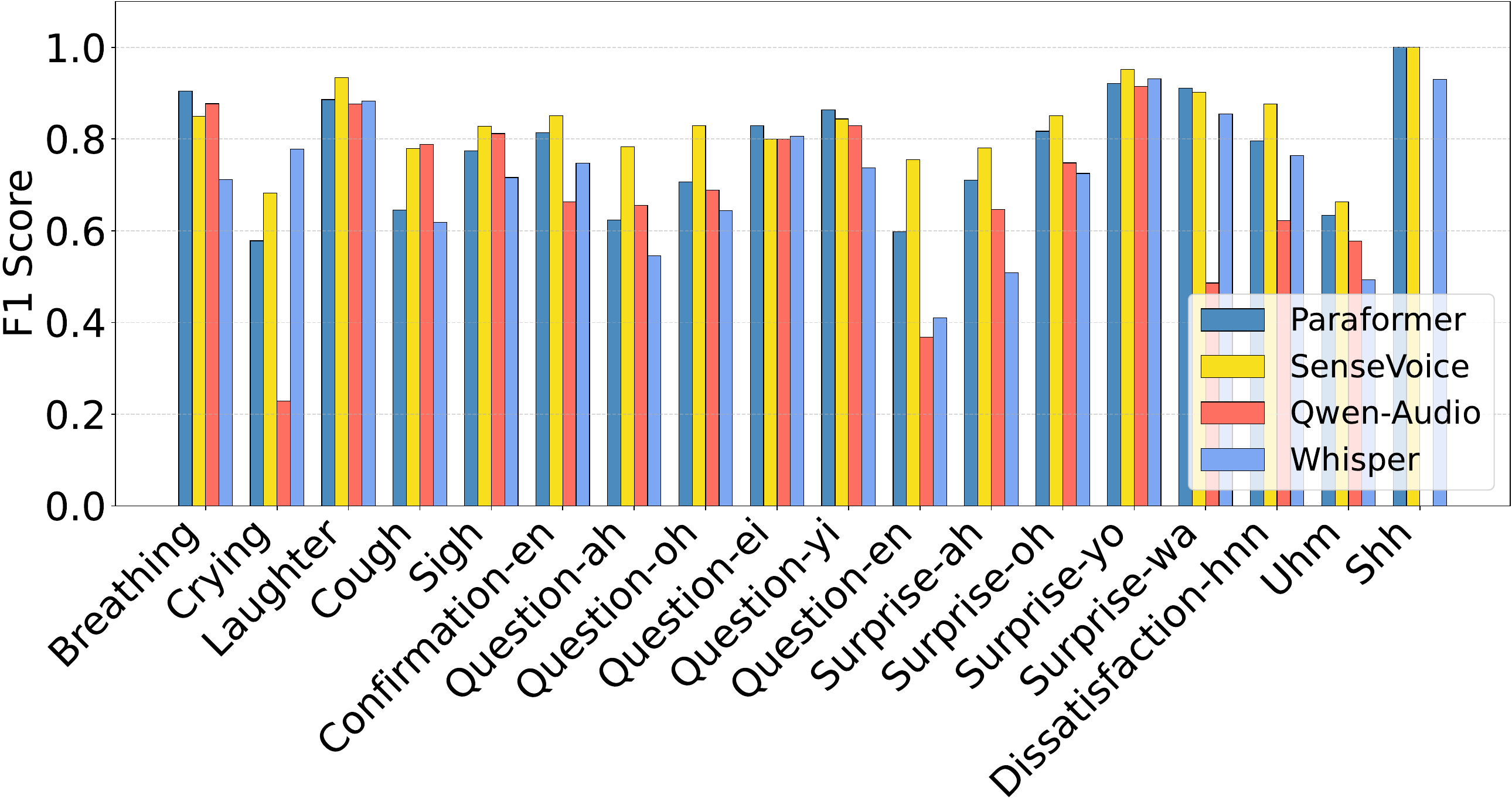}
    \caption{F1 scores across paralinguistic categories.}
    \label{fig:f1_scores_vertical_comparison}
  \end{minipage}
\end{figure}

\subsection{Paralinguistic Aware Speech Recognition}
\label{sec:ASR}
\paragraph{Experimental Setup} We extend conventional ASR to a paralinguistic-aware ASR task by training models to transcribe both lexical content and paralinguistic vocalizations (PV) within a unified token sequence. Each model is provided with paired audio and word-level transcripts, where paralinguistic vocalizations are inserted as special tokens in the target sequence (e.g., ``\texttt{不知道[Breathing]，我没想过}''). This formulation enables the model to treat paralinguistic vocalizations as first-class decoding targets. Specifically, given input audio $x \in \mathbb{R}^{T \times F}$ and target label sequence $y = \{y_1, y_2, ..., y_L\}$, the model learns to minimize the CTC loss~\cite{graves2006connectionist}:
\begin{equation}
\mathcal{L}_{\text{CTC}} = -\log P(y \mid x) = -\sum_{\pi \in \mathcal{B}^{-1}(y)} P(\pi \mid x)
\end{equation}
where $\mathcal{B}$ is the CTC collapsing function and $\pi$ denotes the alignment path. All models are finetuned on the \textbf{NVSpeech\textsubscript{human}} dataset.

\paragraph{Evaluation}
We evaluate ASR models on two test sets. The \textbf{in-domain testset}, a held-out subset of the human-annotated corpus, covers diverse in-game contexts such as greetings, combat, and narrative dialogue. To assess generalization, we also introduce a human-annotated \textbf{open-domain testset} with spontaneous online content including talk shows, interviews, and sports commentary with different accents across different scenarios. We use four metrics for evaluation: CER (character error rate over the full transcript), CER-w/o-para (excluding paralingustic tokens), Para Detection Rate (whether any PV is correctly detected in an utterance), and F1-score (event-level precision and recall on PV prediction). 

\paragraph{Baseline Models} We benchmark four models with distinct architectures and decoding strategies.
\textbf{(1) Paraformer}~\cite{gao2022paraformer} is a non-autoregressive ASR model that employs a continuous integrate-and-fire (CIF) mechanism for segment-wise decoding. Given input audio $x \in \mathbb{R}^{T \times F}$, it first produces hidden representations $h_1, \dots, h_T$, and then uses the CIF gate to compute segmental embeddings:
\begin{equation}
\tilde{h}_i = \text{CIF}(h_1, \dots, h_T)
\end{equation}
We treat PV as special tokens and train the model to emit both lexical and PV labels in a unified output sequence $y = \{y_1, \dots, y_L\}$.

\textbf{(2) SenseVoice-Small}~\cite{an2024funaudiollm} is a non-autoregressive encoder-only model for multi-task speech understanding. The input audio $x \in \mathbb{R}^{T \times F}$ is first converted into 80-dimensional log-Mel filterbanks and then mapped to encoder features $\mathbf{X}_{\text{speech}}$. To specify the ASR task, we prepend a task embedding $\mathbf{e}_{\text{ASR}}$ to the input:
\begin{equation}
\mathbf{X} = \text{concat}(\mathbf{e}_{\text{ASR}}, \mathbf{X}_{\text{speech}})
\end{equation}
The encoder produces contextualized representations, followed by a linear projection and softmax:
\begin{equation}
\mathbf{P} = \text{Softmax}(\text{Linear}_{F \rightarrow |V'|}(\text{Encoder}(\mathbf{X})))
\end{equation}
where $V'$ is the vocabulary including lexical tokens and PV tags. During fine-tuning, we extend the vocabulary with PV-specific tokens and optimize the model using the CTC loss.

\textbf{(3) Qwen-Audio}~\cite{chu2023qwen} combines a Whisper-based audio encoder with a large language model. Although it lacks native support for PV transcription, we extend its output vocabulary with PV-specific tokens and finetune it using instruction-based prompts.

Given a paired input $(x, y)$, where $x \in \mathbb{R}^{T \times F}$ is the input audio and $y = \{y_1, ..., y_L\}$ is the target token sequence including both lexical and PV tokens, the model is trained to maximize the conditional log-likelihood:
\begin{equation}
\label{Eq. 6}
\mathcal{L}_{\text{LM}} = -\sum_{t=1}^{L} \log P_\theta(y_t \mid y_{<t}, \text{Encoder}_\phi(x))
\end{equation}
\textbf{(4) Whisper}~\cite{radford2023robust} is a widely used transformer-based ASR model. Whisper was trained on a very large, diverse dataset covering many languages, accents, and audio conditions. It jointly performs language identification, transcription, and translation. It is optimized with an autoregressive next-token prediction loss, similiar to Eq. \ref{Eq. 6}.

\paragraph{Results}
Table~\ref{tab:para-asr-results} reports results on both in-domain and open-domain testset. On in-domain part, SenseVoice achieves the best CER (4.61\%) and F1-score (0.83), benefiting from its encoder-only SAN-M architecture. Paraformer attains the highest PV Detection Rate (96.1\%), while Qwen-Audio performs worst across all metrics, whose Whisper-initialized encoder and LLM decoder are optimized for semantic abstraction, remains less sensitive to fine-grained paralinguistic cues.
Since in-domain tests mainly reflect performance on game-style speech, we further evaluate on an open-domain set with spontaneous and noisy content to assess robustness in real-world scenarios. Here, SenseVoice again leads (CER 3.79\%, F1 0.85), confirming paralinguistic-aware ASR generalizes effectively beyond controlled domains. Figure~\ref{fig:f1_scores_vertical_comparison} presents detailed F1 scores for all paralinguistic categories. 
The Appendix provides further discussion on model confidence and qualitative analyses of failure cases.

\section{The NVSpeech Dataset}
\subsection{Dataset Construction}
\paragraph{Label Set Definition}
We define 18 categories of paralinguistic vocalizations, encompassing physiological sounds (e.g., \texttt{[Laughter]}, \texttt{[Cough]}), discourse markers (e.g., \texttt{[Uhm]}), and expressive interjections with attitude (e.g., \texttt{[Confirmation-en]}, \texttt{[Question-ah]}, \texttt{[Surprise-oh]}). These labels, derived from empirical analysis of Mandarin spontaneous speech, capture high-frequency paralinguistic sound and functionally distinct categories that support discourse coherence~\cite{tseng2013lexical}.
\paragraph{Data Sources and Augmentation}
We construct the human-annotated dataset from both expressive game-based and augmented sources. The core speech data is drawn from two open-source repositories: the Genshin\footnote{\url{https://huggingface.co/datasets/simon3000/genshin-voice}} and StarRail\footnote{\url{https://github.com/AI-Hobbyist/StarRail_Datasets}} datasets, which offer multilingual voice lines from miHoYo games; we use only the Chinese subset. These lines cover diverse in-game contexts such as greetings, combat, and narrative dialogue, and include metadata such as transcription, speaker identity, and utterance type.

To increase coverage of non-verbal vocalizations, we augment the corpus with 500 coughing and 500 crying clips from Nonspeech7k~\cite{rashid2023nonspeech7k}, a clean, manually labeled non-speech dataset. Additionally, we synthesize 166 utterances using CosyVoice2~\cite{du2024cosyvoice} to enrich rare paralinguistic categories (e.g., \texttt{[Surprise-yo]}, \texttt{[Question-en]}, \texttt{[Shh]}), with text prompts generated via DeepSeek-R1~\cite{deepseekai2025deepseekr1incentivizingreasoningcapability} to ensure contextual diversity and naturalness.

\begin{figure}[h!]
  \centering
  \begin{subfigure}[t]{0.4\textwidth}
    \centering
    \includegraphics[width=0.9\textwidth]{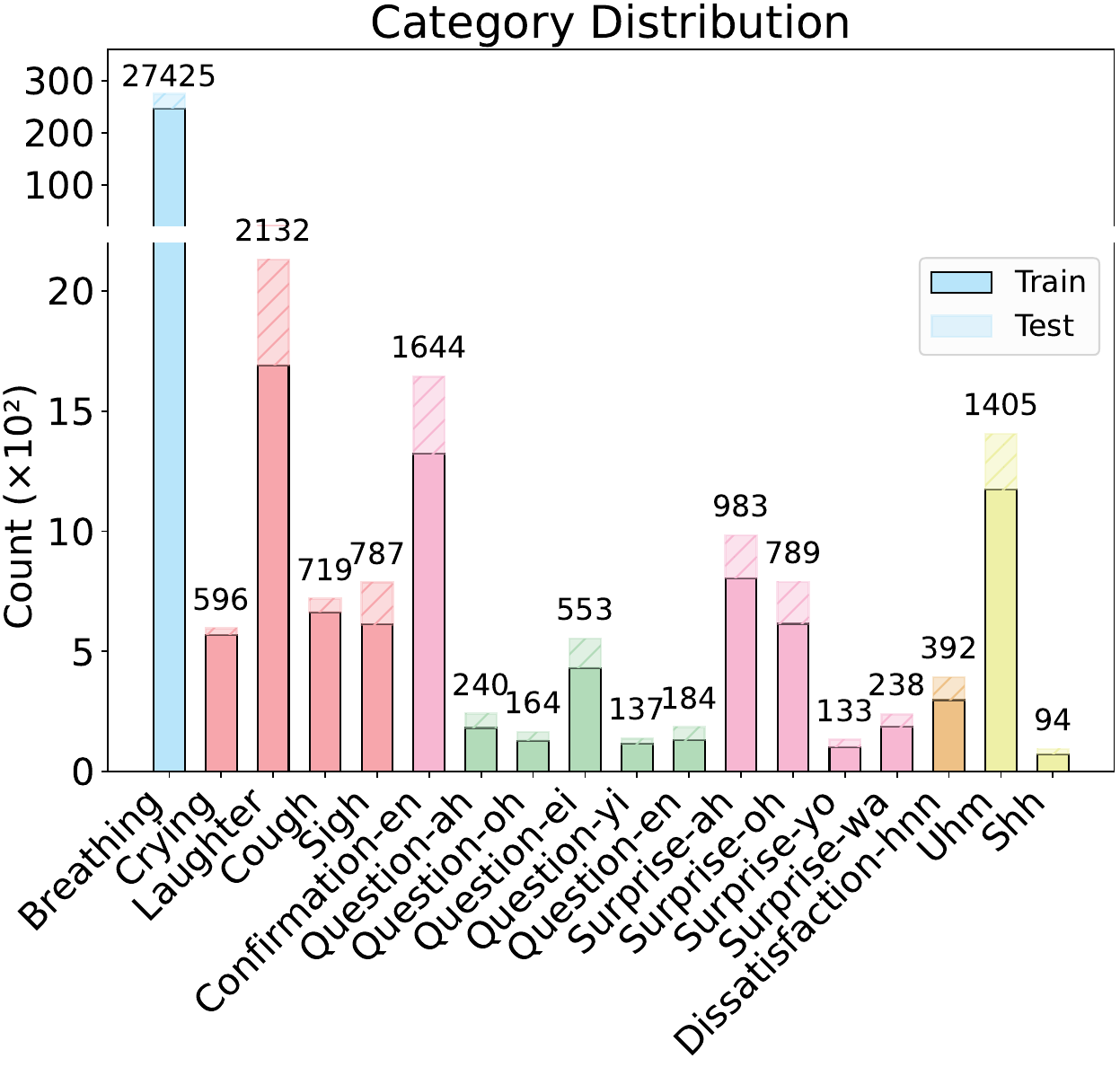}
    \caption{NVSpeech\textsubscript{human}}
    \label{fig:category-distribution}
  \end{subfigure}
  \hfill
  \begin{subfigure}[t]{0.4\textwidth}
    \centering
    \includegraphics[width=0.9\textwidth]{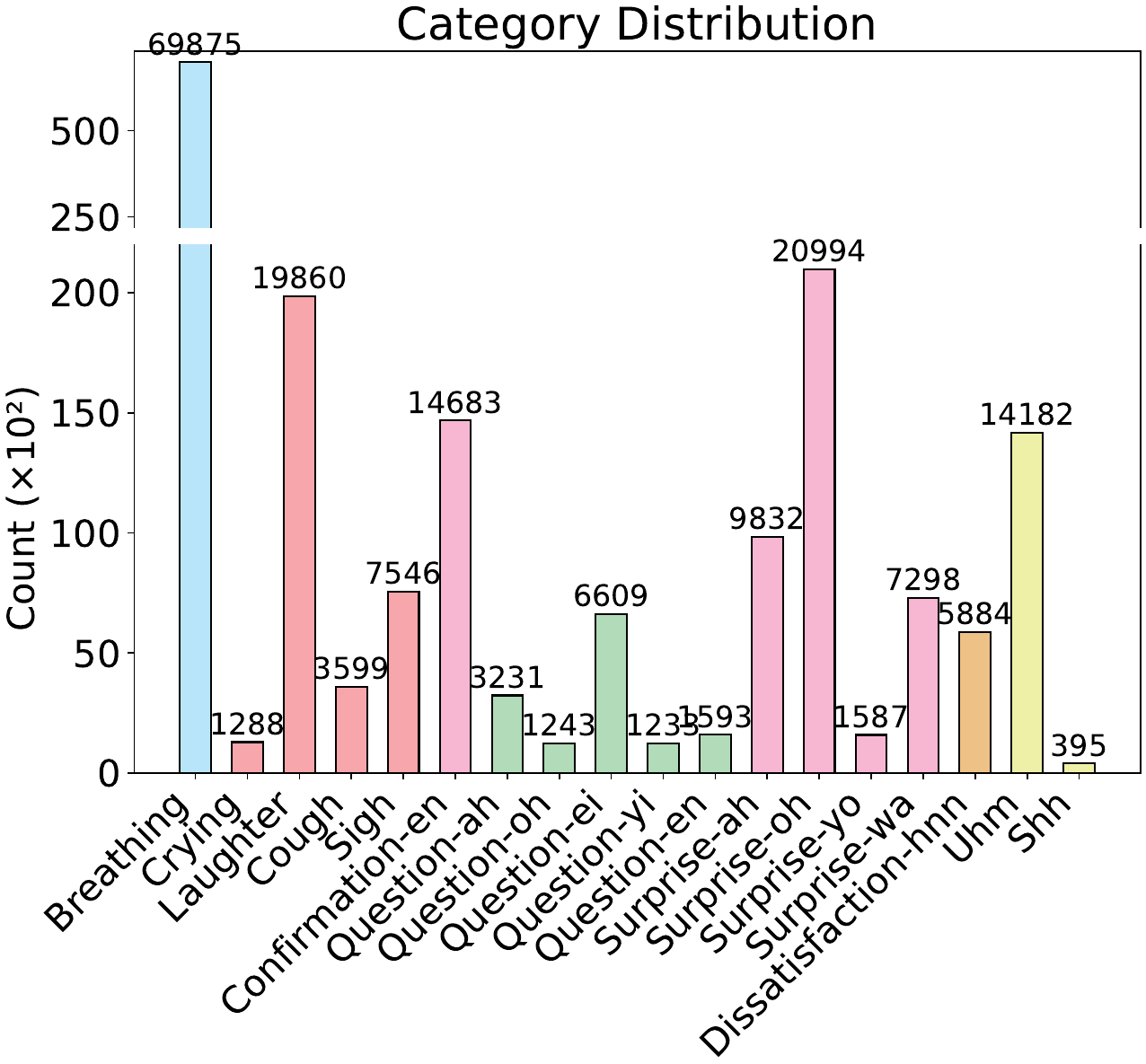}
    \caption{NVSpeech}
    \label{fig:category-autolabel}
  \end{subfigure}
  \caption{Paralinguistic vocalization category distribution}
  \label{fig:overall-nvv-distribution}
\end{figure}

\paragraph{Human-annotated datasets}
Each of the ten trained annotators was tasked with listening to the audio recordings and inserting appropriate paralinguistic vocalization labels into the corresponding transcripts, guided by the temporal location of each label. All annotators received standardized training with positive and negative examples to ensure consistency. For quality assurance, 5\% of the data was cross-annotated, yielding a Cohen’s kappa above 0.85 on major categories. The overall annotation workflow is illustrated in Step1 of Figure \ref{fig:enter-label}, and the label distribution is illustrated in Figure~\ref{fig:category-distribution}.

\subsection{Large-scale automatically scaled datasets}
To further scale our training corpus beyond the manually labeled subset, we construct a large-scale automatically annotated dataset using high-quality speech data from multiple sources. Specifically, we include: (1) the unlabeled portion of the Genshin and StarRail dataset (excluding the manually annotated subset), 
(2)A subset of Emilia\cite{he2025emilia}, a large-scale multilingual speech dataset constructed from in-the-wild recordings, including talk shows, interviews, debates, and audiobooks. We select clips likely to contain paralinguistic events.
(3) 1,362 non-verbal clips (\texttt{[Crying]} and \texttt{[Cough]}) sampled from Nonspeech7k~\cite{rashid2023nonspeech7k}. All clips are in Chinese and capture rich expressive behaviors from diverse sources, ranging from structured in-game to spontaneous, real-world conversational scenarios.

We use SenseVoice, the best-performing paralinguistic-aware ASR (Section~\textit{Paralinguistic Aware Speech Recognition}), to automatically transcribe audio with both lexical content and inline tags for paralinguistic vocalizations.

In total, the automatically labeled dataset contains 174,179 audio-transcription pairs, amounting to 573.4 hours, significantly expanding the diversity and coverage of paralinguistic vocalization categories. Figure~\ref{fig:category-autolabel} shows the label distribution, and Table~\ref{tab:para-datasets} compares it with existing paralinguistic datasets. This large-scale dataset serves as a valuable resource for pretraining and semi-supervised learning, reducing the reliance on costly human annotation. 

\begin{table*}[t!]
  \centering
  \vspace{-1em}
  \renewcommand{\arraystretch}{1.2}
  \begin{tabular}{lcccc|cccc}
    \toprule
    \textbf{Model} &
    \multicolumn{4}{c|}{\textbf{In-domain Testset}} &
    \multicolumn{4}{c}{\textbf{Open-domain Testset}} \\
    \cmidrule(lr){2-5} \cmidrule(lr){6-9}
    & \textbf{CER$\downarrow$} & \textbf{CER$_{\text{w/o para}}$$\downarrow$} & \textbf{SIM$\uparrow$} & \textbf{UTMOS$\uparrow$}
    & \textbf{CER$\downarrow$} & \textbf{CER$_{\text{w/o para}}$$\downarrow$} & \textbf{SIM$\uparrow$} & \textbf{UTMOS$\uparrow$} \\
    \midrule
    \rowcolor{gray!15}
    \multicolumn{9}{c}{\textit{Pre-trained}} \\
    CosyVoice          & - & 7.42\% & 0.727 & \textbf{2.69} & - & 10.44\% & 0.743 & \textbf{2.49}  \\
    CosyVoice2         & - & \textbf{3.13\%} & 0.710 & \textbf{2.69} & - & 7.91\% & 0.722 & 2.25  \\

    \rowcolor{gray!15}
    \multicolumn{9}{c}{\textit{Finetuned on Human-Labeled Data}} \\
    CosyVoice          & 8.78\% & 4.21\% & \textbf{0.736} & 2.54 & 11.09\% & 6.71\% & \underline{0.748} & 2.35  \\
    CosyVoice2         & 8.61\% & 3.86\% & 0.709 & 2.54 & 9.48\% & \underline{5.57\%} & 0.719 & 2.12  \\

    \rowcolor{gray!15}
    \multicolumn{9}{c}{\textit{Finetuned on Auto-Labeled Data (Equal Size)}} \\
    CosyVoice          & 8.59\% & 4.07\% & \textbf{0.736} & 2.54 & 9.97\% & 6.12\% & \textbf{0.750} & 2.35  \\
    CosyVoice2         & \underline{7.83\%} & 3.77\% & 0.704 & 2.57 & \underline{8.44\%} & \textbf{5.45\%} & 0.710 & 2.20 \\

    \rowcolor{gray!15}
    \multicolumn{9}{c}{\textit{Finetuned on Auto-Labeled Data (Large-Scale)}} \\
    CosyVoice          & 7.96\% & 4.05\% & \underline{0.733} & 2.57 & 9.99\% & 5.84\% & 0.747 & \underline{2.39}  \\
    CosyVoice2         & \textbf{7.51\%} & \underline{3.73\%} & 0.700 & \underline{2.67} & \textbf{8.07\%} & 5.73\% & 0.703 & 2.26 \\
    \bottomrule
  \end{tabular}
  \caption{Objective Evaluation of Para-enhanced TTS. \textbf{Bold} indicates best in the column, \underline{underline} second-best.}
  \label{tab:tts-objective-full}
\end{table*}

\section{Paralinguistic-enhanced TTS Experiments}
In this section, we evaluate the effectiveness of the proposed \textbf{NVSpeech} dataset in training zero-shot TTS capable of generating expressive speech with natural paralinguistic behaviors.
\subsection{Experimental Setups}
To evaluate the effectiveness of the automatically labeled \textbf{NVSpeech} dataset, we conduct TTS enhancement experiments on three training subsets: (1) \textbf{NVSpeech\textsubscript{human}} (human-annotated), (2) \textbf{NVSpeech\textsubscript{human-size}} (an auto-labeled subset with the same size, preserving a similar label distribution as NVSpeech\textsubscript{human}), and (3) \textbf{NVSpeech} (full auto-labeled). We finetune a pretrained TTS model, extending their vocabularies to include paralinguistic tags. Training data consists of 35\% regular speech and 65\% paralinguistic-rich utterances.

\paragraph{Baseline Models}
CosyVoice~\cite{an2024funaudiollm} is a zero-shot TTS system that leverages supervised semantic tokens, using a language model to predict tokens from text and a flow-matching decoder to synthesize speech.
CosyVoice2~\cite{du2024cosyvoice} is an instruction-following TTS model that supports paralinguistic prompts; we extend its vocabulary to include NVSpeech tags for fine-grained control over event placement. 

\paragraph{Evaluation}
We evaluate TTS models on both in-domain and open-domain testsets. The in-domain testset, drawn from the human-annotated corpus, reflects performance on controlled game-style speech. The open-domain testset with spontaneous and diverse real-world speech, which better reflects practical application scenarios and show robustness beyond in-domain testset.
Objective metrics include overall character error rate (CER), CER on verbal content only without paralingustic labels (CER\_wo\_para), speaker similarity (SIM)~\cite{san2017simplified}, and UTMOS~\cite{saeki2022utmos} for perceptual audio quality. 

\subsection{Main Results}

Table~\ref{tab:tts-objective-full} summarizes the objective evaluation of para-enhanced TTS.
(1) \textbf{Human-labeled finetuning} enables paralinguistic generation while slightly lowering CER overall, with no consistent degradation across models or domains.
(2) \textbf{Auto-labeled data (equal size)} yields better CER and UTMOS than human-only training with comparable SIM, showing that auto-labeled data can match the effectiveness of human annotation. 
(3) \textbf{Scaling with large auto-labeled data} achieves the best performance, with up to 12.8\% relative CER reduction on in-domain speech.

These results verify the effectiveness of our pipeline for both game-related and open-domain TTS tasks, and demonstrate its scalability—showing that both large-scale auto-annotation and human annotation can substantially enhance paralinguistic synthesis.

\subsection{Human Evaluation}
\paragraph{Evaluation Metrics}
Subjective evaluation considers human preference scores, paralinguistic event recall, perceived naturalness of events synthesis and transitions (NMOS), and QMOS for overall quality of synthesized speech.

\begin{figure}[h]
    \centering
    \includegraphics[width=\linewidth]{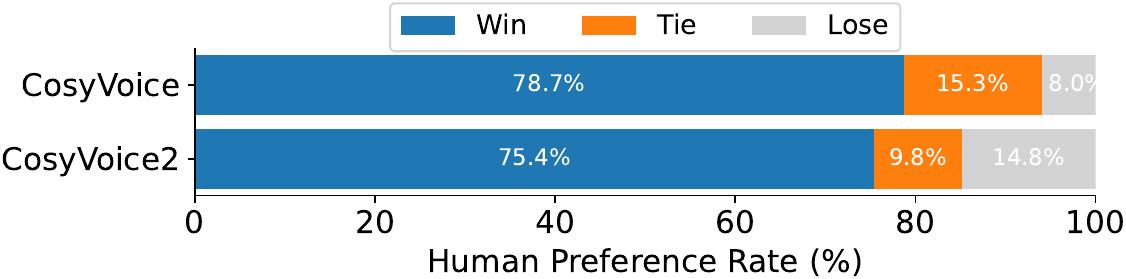}
    \caption{Para-enhanced vs. Original (Human Preference).}
    \label{fig:win_tie_lose_barh}
\end{figure}

We invited 60 participants to compare TTS outputs before and after fine-tuning with \textbf{NVSpeech}. As shown in Figure~\ref{fig:win_tie_lose_barh}, both CosyVoice and CosyVoice2 saw clear listener preference after para-enhancement, with win rates of 78.7\% and 75.4\%, respectively. Table~\ref{tab:Subjective Evaluation} further shows that the finetuned models achieved high naturalness (NMOS: 3.9-4.0) and clarity (QMOS: 4.04-3.96), while maintaining reasonable recall of paralinguistic tags (up to 61.9\%). These results confirm that \textbf{NVSpeech} enables natural and expressive speech synthesis with structured paralinguistic cues, without compromising core speech quality.

\begin{table}[h!]
    \centering
    \begin{tabular}{cccc}
        \toprule
        \textbf{Model} & \textbf{Recall$\uparrow$} & \textbf{NMOS$\uparrow$} & \textbf{QMOS$\uparrow$} \\
        \midrule
        CosyVoice & 0.604 & 3.9 ± 0.20 & \textbf{4.04 ± 0.15} \\
        CosyVoice2 & \textbf{0.619} & \textbf{4.0 ± 0.16} & 3.96 ± 0.14 \\
        \bottomrule
    \end{tabular}
    \caption{Subjective Evaluation of Para-enhanced Speech.}
    \label{tab:Subjective Evaluation}
\end{table}

\section{Conclusion} 
We present NVSpeech, an integrated and scalable pipeline for paralinguistic-aware speech dataset, recognition, and generation.
The pipeline first establishes a word‑level paralinguistic resource with 18 paralinguistic vocalization categories, then trains a paralinguistic‑aware ASR on the human‑annotated subset to automatically label large‑scale data (573.4h), and finally enhances zero‑shot TTS with both human‑ and auto‑labeled data. Experimental results demonstrate strong paralinguistic tag recognition (F1 up to 0.84) and expressive speech synthesis preferred by listeners (win rate 78.7\%) without degrading lexical quality, confirming the pipeline’s effectiveness for both recognition and generation. This work provides a scalable foundation for future research on expressive, human‑like speech modeling.

\bibliography{aaai2026}

\clearpage


\section*{Supplementary material}

# Supplementary Materials

This supplementary material for paper, "NVSpeech: An Integrated and Scalable Pipeline for Human-Like Speech Modeling with Paralinguistic Vocalizations" describes the details of technical aspects.

## A   Experiment Details

**PANNs [7]** We performed end-to-end fine-tuning of the `Wavegram_Logmel_Cnn14` model using the PANNs-provided script on four NVIDIA RTX 4090. Training was carried out for 50,000 iterations with a batchsize of 32 and a learning rate of 1e-4, employing the AdamW optimizer, and consumed approximately 22 GB of GPU memory.

**SenseVoice-Small [1]** We performed end-to-end finetuning of the SenseVoice-Small model on two NVIDIA RTX 4090 GPUs. Training was carried out for 50 epochs with dynamic batchsize and a learning rate of 1e-4 for paralinguistic tagging, 4e-5 for paralinguistic ASR to get the best performance. We employing the AdamW optimizer, and consumed approximately 21 GB of GPU memory.

**Paraformer [5]** We performed end-to-end finetuning of the Paraformer model on eight NVIDIA RTX 4090 GPUs. Training was carried out for 100 epochs with dynamic batchsize and a learning rate of 5e-4. We employing the AdamW optimizer, and consumed approximately 21 GB of GPU memory.

**Qwen-Audio [3]** We performed LoRA strategy to finetune Qwen-Audio. We employ the AdamW optimizer with 2e-4 learning rate for 5 epochs. Training was carried out with four batchsize on eight A100 80GB and consumed approximately 70GB of GPU memory.

**Whisper [8]** We performed end-to-end finetuning of the Whisper. We employ the AdamW optimizer with 1e-5 learning rate for 5 epochs. Training was carried out with four batchsize on eight A100 80GB.

**CosyVoice [1]** We used official finetune script to finetune CosyVoice LLM model on four A100 80GB. Training was carried out for 30 epochs with dynamic batchsize and a learning rate of 1e-5. We employing the Adam optimizer

**CosyVoice2 [4]** We used official finetune script to finetune CosyVoice2 LLM model on four A100 80GB. Training was carried out for 20 epochs with dynamic batchsize and a learning rate of 1e-5. We employing the Adam optimizer, and consumed approximately 35 GB of GPU memory.

## B   Confusion Patterns Analysis of Paralinguistic-Aware ASR

### B.1   Confusion Matrix Analysis of paralinguistic-aware ASR

To further assess the recognition performance of our paralinguistic-aware ASR systems, we present confusion matrices for the four evaluated models on the NVSpeech human-labeled testset (Figure 1). Each matrix shows prediction counts per ground-truth label, with intensity proportional to frequency.

All four models achieve high accuracy on frequent events such as [Breathing] and [Laughter], with Paraformer (3785, 399) and SenseVoice-Small (3110, 402) leading in detection counts. SenseVoice-Small also maintains balanced recall across mid-frequency events such as [Cough] (37) and [Question-ah] (46), indicating robustness in varied conversational contexts. Whisper excels in precision for dominant categories, making it suitable for applications prioritizing common-event accuracy, while Qwen-Audio retains competitive performance on high-impact categories and benefits from multi-modal training for cross-domain adaptability.

**Confusion Patterns** (1) High-frequency categories – Confusions mainly arise from acoustic similarity, e.g. [Breathing] vs. [Sigh]. (2) Mid-frequency categories – Confusions stem from similar prosodic patterns, e.g. [Confirmation-en] vs. [Uhm]. (3) Rare categories – Confusions are driven by data sparsity and subtle pitch/timbre differences, e.g. the four [Surprise-*] tags.

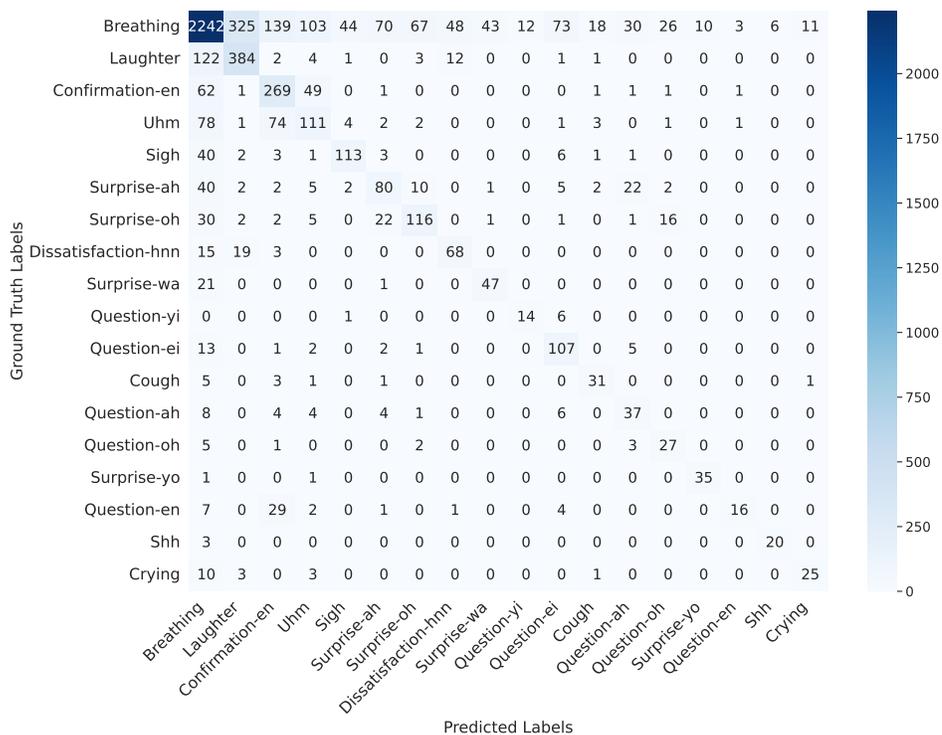

(a) Whisper

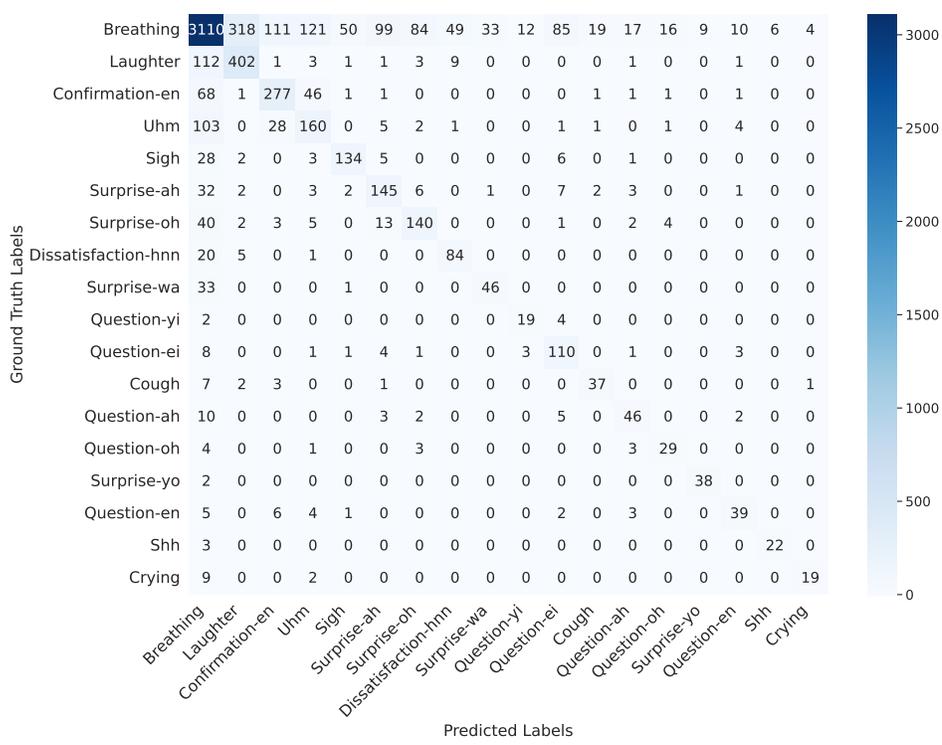

(b) SenseVoice-Small

Figure 1: Confusion matrices of four paralinguistic-aware ASR models (Part 1/2).

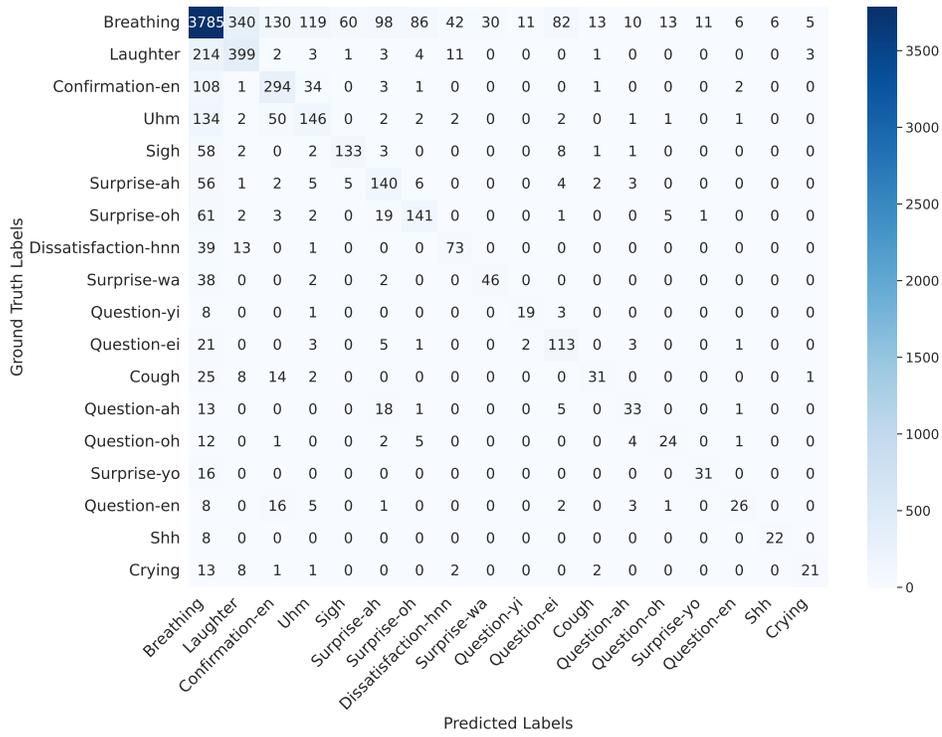

(c) Paraformer

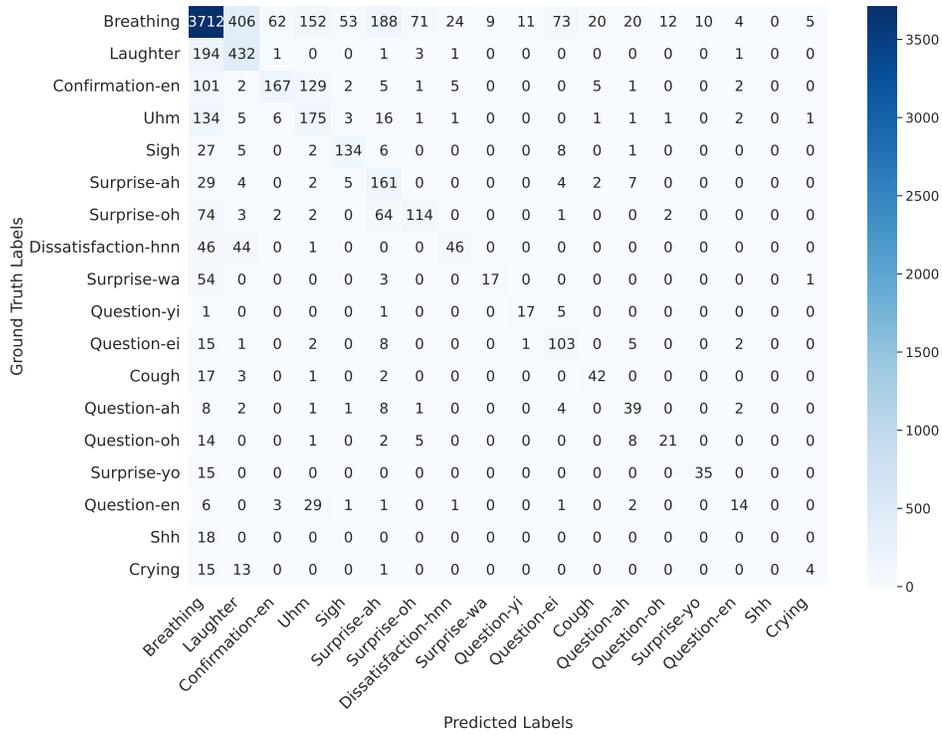

(d) Qwen-Audio

Figure 1: Confusion matrices of four paralinguistic-aware ASR models (Part 2/2)

3

## B.2 Confidence Analysis

**Top-1 vs. Top-2 Predicted Label Confidence Dynamics** To better understand the paralinguistic-aware ASR model's decision confidence for correctly predicted paralinguistic events, we plot the top-1 and top-2 confidence scores of each paralinguistic events on both the in-domain and open-domain test sets in Figure 2.

We observe that points are significantly denser in the lower-right region (high top-1 confidence and low top-2 confidence), indicating that the model often makes confident and unambiguous predictions. This is especially prominent for well-separated classes such as [Crying] and [Question-yi], where alternative labels are assigned substantially lower probabilities.

In contrast, a smaller subset of points appears in regions where both top-1 and top-2 confidence are relatively low and close in value (e.g. top-1 $< 0.6$ and top-2 $> 0.3$), reflecting more ambiguous cases. For instance, when [Question-en] is predicted with top-1 confidence, top-2 candidates often include [Confirmation-en] and [Uhm], suggesting the model is sensitive to prosodic similarities among these functionally related cues. These patterns highlight opportunities for refinement through confidence-aware training or targeted sample enhancement.

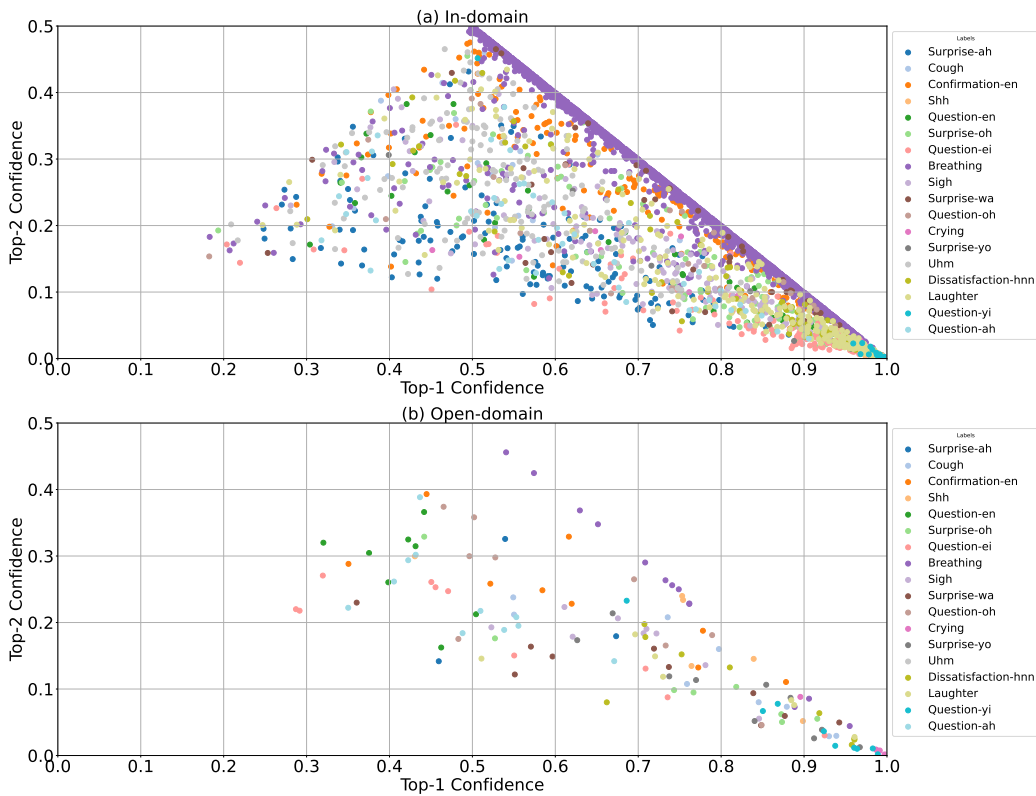

Figure 2: Top-1 vs. Top-2 confidence scores for correctly predicted paralinguistic events.

**Confidence Distribution Analysis** As shown in Figure 3 and Figure 4, the model exhibits distinct top-1 confidence patterns across paralinguistic labels for correct predictions under in-domain and open-domain settings. In both cases, categories such as [Crying], [Question-yi], [Surprise-yo], and [Dissatisfaction-hnn] demonstrate high and consistent confidence (mean $> 0.8$), indicating strong model certainty and clear acoustic discrimination.

Moderate confidence levels (0.7–0.8) are observed for labels like [Laughter], [Cough], and [Surprise-oh], suggesting stable yet more variable acoustic realizations. In contrast, lower and more dispersed confidence distributions are found for [Question-en], [Question-ah], and [Surprise-ah], particularly in the open-domain testset (Figure 4). This highlights the model's challenges in generalizing to acoustically ambiguous or prosodically similar cues, emphasizing the need for confidence-aware calibration or targeted augmentation.

4

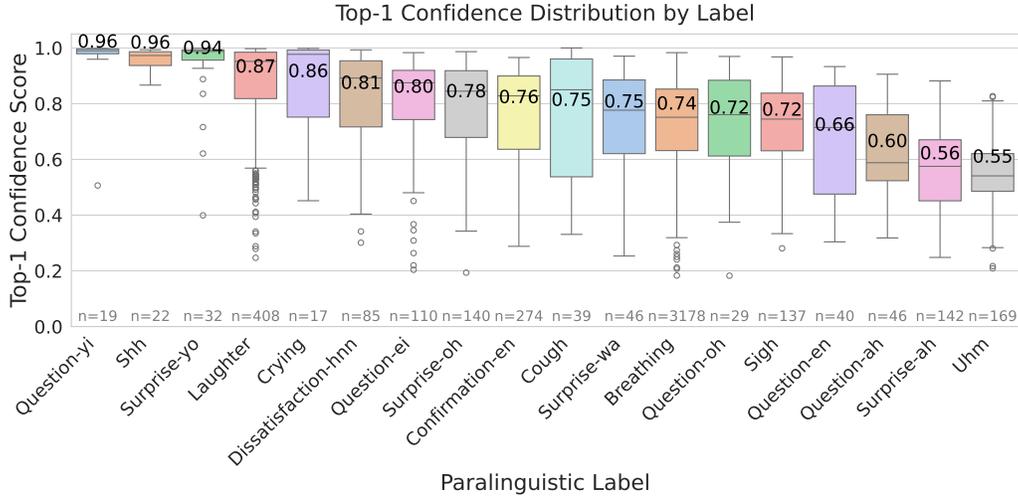

Figure 3: Top-1 confidence distribution per label for correct predictions (In-domain).

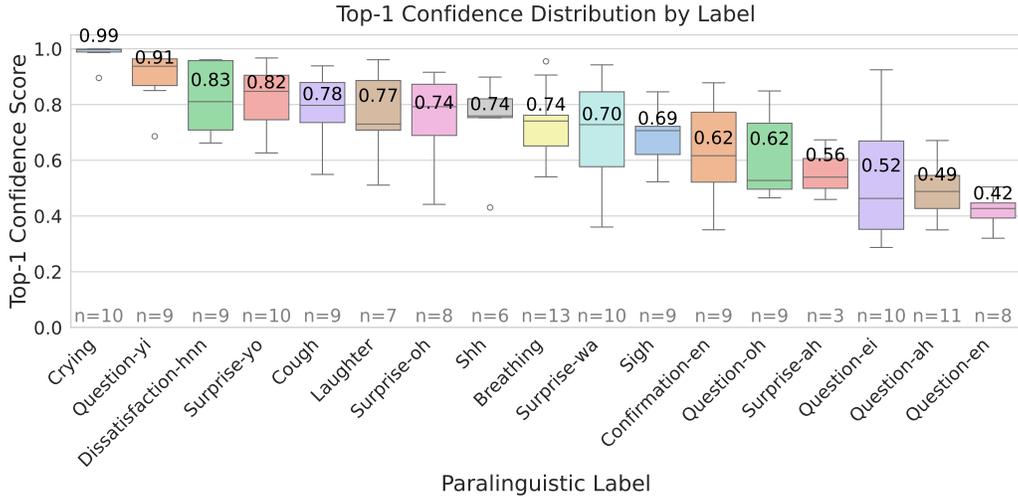

Figure 4: Top-1 confidence distribution per label for correct predictions (Open-domain).

## C  NVSpeech Dataset Analysis

To enable a comprehensive stress-test of our ASR model, the open-domain testset is intentionally composed of hard cases. It targets multiple independent axes of difficulty so that weaknesses in robustness, temporal modeling, domain generalization, and entity fidelity are exposed. Table 1 presents representative hard-case examples and Figure 5 shows the distribution of speaking rate and audio duration.

1. **Spontaneous Disfluencies:** Includes repetitions and self-repairs (e.g., "not me me me," "I can't control it, I can't control it"), probing the model's resilience to realistic conversational hesitation and inline corrections.

2. **Proper Nouns:** Terms that uniquely identify specific entities such as "Qin Shi Huang".

3. **Names:** Personal or historical names, such as "Gongzi Ang," or the direct Chinese transliterations of "Joe" and "Aamon."

4. **Idioms:** Fixed expressions conveying cultural or historical meaning, e.g., "位极人臣" (to reach the pinnacle of ministerial office).

5

5. **Source Diversity:** Audio is drawn from a wide range of domains—talk shows, interviews, sports commentary, e-books, and real-life recordings—to evaluate generalization across registers, styles, and acoustic conditions. Some audio even contain noticeable background noise.

6. **Speaking Rate Distribution:** Slow ($\leq 3.00$ chars/sec): 40 utterances (25.5%); medium (3.00–4.50 chars/sec): 94 utterances (59.9%); fast ($> 4.50$ chars/sec): 23 utterances (14.6%), challenging temporal adaptation and alignment.

7. **Audio Duration Variety:** Short ($\leq 5$ seconds): 27 clip (17.2%); medium (5–15 seconds): 91 clips (58%); long ($> 15$ seconds): 39 clips (24.8%); see Figure 5 for the distribution.

By probing these orthogonal dimensions, we can evaluate the model's robustness and fidelity across diverse out-of-domain scenarios.

| Type | Target |
|---|---|
| Disfluency | 不是我我我，就是我、就是我没法管 |
| Proper noun | 这时刘邦便想起来秦始皇 |
| Name | 乔伊对亚蒙这个能干又温柔的助理还挺有好感的 |
| Idiom | 我说[Dissatisfaction-hnn]敬酒不吃吃罚酒 |

Table 1: Four types of hard cases in open-domain testset

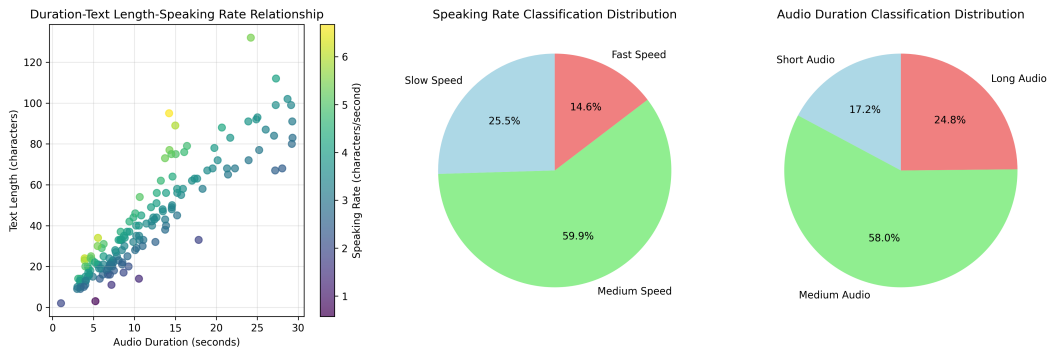

Figure 5: The distribution of speaking rate and audio-length. Slow speed ($\leq 3.00$ chars/sec), medium speed (3.00–4.50 chars/sec), fast speed ($> 4.50$ chars/sec). Short audio ($\leq 5$ seconds), Medium audio (5–15 seconds), Long audio ($> 15$ seconds).

## D  Paralingustic-aware English ASR

**Experimental Setup**  We extend paralingustic-aware ASR on English datasets to prove the effectiveness of paralingustic-aware ASR in multiple languages. Similarly, each model is provided with paired audio and word-level transcripts, where paralinguistic vocalizations are inserted as special tokens in the target sequence (e.g., "[Uhm] I was so impressed with that movie").

**Datasets**  We employ two open-source English datasets with paralingustic cues. **DisfluencySpeech** [9], a studio-quality labeled English speech dataset with paralanguage. It holds 9.49 hours of single speaker speech, about 5000 utterances from Switchboard-1 Telephone Speech Corpus (Switchboard). It focus on disfluencies and filled pauses in the daily speech, which can be categorized into five types: filled pauses (e.g. "uh"), editing terms (e.g. "I mean"), discourse markers (e.g. "you know"), coordinating conjunctions (e.g. "and", "but"), asides, restarts and non-speech sounds (e.g. "Laughter"). **NonverbalTTS** [2], a 17 hours English speech dataset, from VoxCeleb and Expresso, with 10 categories of nonverbal vocalizations (NVVs) (e.g. Laughter) and 8 emotional categories.

6

**Label Conversion** To make the English datasets compatible with our existing Chinese paralinguistic cues, we project and normalize their NVVs annotations into a shared label space. We extract utterances containing five target NVVs—[Uhm], [Cough],[Laughter], [Sigh], and [Breathing]—and also include ordinary speech segments with no NVVs [None] as negative examples from these two datasets. Non-matching NVVs are filtered out unless they semantically overlap (e.g., certain subtle inhalation or hesitation sounds mapped to [Breathing] or [Uhm] based on acoustic similarity). All special NVVs are inserted into the target transcript as discrete tokens (e.g., "[Laughter] That was amazing") so the model learns to jointly predict lexical content and paralinguistic events.

**Dataset Splits and Statistics** The consolidated English paralinguistic-aware dataset comprises 9,457 utterances, partitioned into disjoint splits: 8,412 utterances for training, 101 for validation, and 944 for testing. Table 2 summarizes the split sizes and NVVs prevalence.

| Split | Utterances | None | Breathing | Laughter | Cough | Sigh | Uhm |
|---|---|---|---|---|---|---|---|
| Train | 8,412 | 3,136 | 3,184 | 1,455 | 426 | 153 | 1,753 |
| Validation | 101 | 15 | 47 | 17 | 9 | 1 | 16 |
| Test | 944 | 400 | 316 | 106 | 36 | 3 | 206 |

Table 2: English paralinguistic-aware dataset composition after label conversion.

**Models and Baselines** The baseline models are similar to those used in the Chinese paralinguistic-aware ASR described in the main content. **(1) SenseVoice-Small** [1] is a multilingual ASR model for multi-task speech understanding. **(2) Qwen-Audio** [3] combines a Whisper-based audio encoder with a large language model. **(3) Whisper** [8] is a multilingual, transformer-based automatic speech recognition model pre-trained with large-scale weak supervision. However, we do not evaluate Paraformer [6] on the English testset because it only supports Chinese speech.

**Results** The results reveal a trade-off between lexical accuracy and paralinguistic sensitivity; this is compounded by dataset difficulty—multiple speakers with varied accents and DisfluencySpeech's repetitions, restarts, and half-articulated words make transcription and NVVs detection harder.

Whisper achieves the lowest overall CER (10.0%) and the best "no-para" baseline (7.29%), indicating the strongest pure lexical modeling. The degradation when inserting paralinguistic tokens suggests a non-trivial interference, but its pretraining gives it enough capacity to absorb much of that cost. However, its paralinguistic detection is comparatively weaker (79.1% detection rate, F1 0.79), implying it is less sensitive or precise in identifying and localizing NVVs even while preserving word-level fidelity.

SenseVoice strikes a more balanced compromise. Its CER is only slightly higher (10.37%), but the relative impact of modeling paralinguistic content is smaller, meaning its joint modeling disturbs the base transcription less. More importantly, it leads on the paralinguistic side with the highest detection rate (88.9%) and F1 (0.84), showing superior ability to capture NVVs accurately. This makes SenseVoice preferable in applications where both the words and the accompanying paralinguistic signals matter and must be retained in tandem.

Qwen-Audio underperforms on both fronts: its high CER (17.76%) and weaker paralinguistic metrics suggest either a mismatch in its capacity for this English NVVs-augmented setting or deficiencies in how it represents or integrates paralinguistic cues.

The results show that paralinguistic-aware ASR models can transcribe paralinguistic events across languages, demonstrating their broad effectiveness and feasibility.

| Model | CER↓ | CER$_{\text{w/o para}}$↓ | Para Det. Rate↑ | F1-score↑ |
|---|---|---|---|---|
| Whisper | **10.06%** | **7.29%** | 79.1% | 0.79 |
| SenseVoice | 10.37% | 8.94% | **88.9%** | **0.84** |
| Qwen-Audio | 17.76% | 15.54% | 83.2% | 0.80 |

Table 3: Performance of paralinguistic-aware ASR on the English testset.

# E List of Paralingustic labels

The list of paralingustic labels is shown in Table 4. These labels are organized into three categories: non-verbal vocalizations, prosodic and attitudinal cues, and discourse-like markers. Each label denotes either a non-verbal event, an emotional cues, or a specific discourse-like marker. Labels in the non-verbal vocalization and discourse-like marker categories are named according to the corresponding audio event, while those in the prosodic and attitudinal cue category are labeled using the associated Chinese phoneme combined with the relevant emotional intent.

| Label | Description |
| --- | --- |
| Breathing | The sound of human inhalation and exhalation. |
| Crying | The sound of human distress with intermittent sobs and heaving breaths. |
| Sigh | A long, audible exhalation conveying resignation, fatigue. |
| Laughter | The sound of human amusement with intermittent vocalizations. |
| Cough | A sudden, forceful expulsion of air from the lungs. |
| Uhm | A brief, voiced hesitation marker. |
| Shh | A soft sound used to urge silence. |
| Surprise-ah | A sudden, sharp exclamation "ah" expressing surprise. |
| Question-ah | A brief "ah" with a falling intonation, used to signal a question. |
| Surprise-oh | A sudden, sharp exclamation "oh" expressing surprise. |
| Question-oh | A brief vocalization "oh" with a falling intonation. |
| Confirmation-en | A brief voiced interjection "en," used to signal acknowledgment. |
| Question-en | A brief vocalization "en" with a falling intonation. |
| Dissatisfaction-hnn | A brief "hnn" interjection conveying annoyance or dissatisfaction. |
| Question-ei | A brief vocalization "ei" with a falling intonation. |
| Question-yi | A brief vocalization "yi" with a falling intonation. |
| Surprise-yo | A sudden, sharp exclamation "yo" expressing surprise. |
| Surprise-wa | A sudden, breathy exclamation "wa" expressing surprise. |

Table 4: The definition of paralingustic labels

# F Annotation User Interfaces

As shown in Figure 6, the annotation user interface presents users with an audio clip and its corresponding transcription. Users can select appropriate paralingustic labels from a predefined label set and insert them into specific positions within the transcription, resulting in an enriched transcript that includes paralingustic vocalizations.

Figure 6: Annotation User Interfaces

## G  Human Evaluation

Below, we provide an example of our human evaluation.

# Audio Evaluation

**Task Description (Must Read Before Answering)**

- **Target:** Each item presents two audio samples. Please respond to the questions outlined below, assigning a score to each audio sample and indicating your preferred sample.
- **Evaluation Criteria:**
    1. **Quality MOS (QMOS)** — Evaluate audio fidelity (any unnatural noise, artifacts, electric buzz, or screeching).
    2. **Naturalness MOS (NMOS)** — Evaluate how natural the speech sounds, including non-verbal vocalizations cues and the fluency between those cues and the main text.
- **Priority:** First consider whether non-verbal vocalizations from the target text is correctly rendered; then check that basic TTS performance has not degraded.

**Scoring Definitions**

**Naturalness MOS (NMOS), 1–5**

- 1: Clearly robotic or distorted; non-verbal vocalizations cues feel forced.

- 2: Audible stiffness; minor discontinuities in phrasing.
- 3: Occasional synthetic tone; non-verbal vocalizations cues mostly natural; generally coherent.
- 4: High naturalness with only brief synthetic artifacts; seamless transition between cues and speech.
- 5: Indistinguishable from human; prosody, emotion, and cues perfectly integrated.

**Quality MOS (QMOS), 1–5**

- 1: Severe noise, clipping or jitter hindering comprehension.
- 2: Slight distortion or background hush; occasional harshness.
- 3: Minor, infrequent noise/distortion; overall clear.
- 4: Very clean; only barely perceptible compression artifacts.
- 5: Pristine, with no noise or distortion; professional-grade quality.

**Audio Samples:**

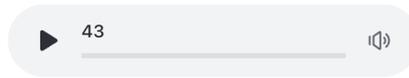
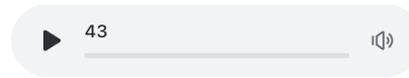

Audio A　　　　　　　　　　　　　　Audio B

**Questionnaire:**

For Audio A, assign:

- *NMOS* (Naturalness MOS): _____
- *QMOS* (Quality MOS): _____

For Audio B, assign:

- *NMOS* (Naturalness MOS): _____
- *QMOS* (Quality MOS): _____

Then select which audio sample you prefer::

| Audio A | Audio B | Equivalent Quality |

## H  Limitation

Our dataset, NVSpeech, which is the largest dataset annotated with word-level paralinguistic vocalizations", providing a foundation for training TTS models capable of emulating human-like non-verbal vocal behaviors, contributes to a more human-like audio synthesis system. However, it may also introduce some social risks and shortcomings. Enhanced verisimilitude in TTS synthesis markedly diminishes the ability to distinguish authentic from synthetic speech, thereby eroding both interpersonal and institutional trust. By replicating the vocal characteristics of real individuals with high fidelity, such systems facilitate the surreptitious distribution of misinformation and audio-based deepfakes, which can distort public discourse, compromise legal proceedings, and imperil personal security. This ambiguity in speech provenance intensifies challenges in the forensic verification of audio evidence.



\clearpage\end{CJK*} 
\end{document}